\documentclass[aps,pre0,twocolumn]{revtex4-2}
\usepackage[utf8]{inputenc}
\usepackage{bm}
\usepackage{amssymb}
\usepackage{mathtools}
\usepackage{amsmath}
\usepackage{xcolor}
\usepackage{enumitem}
\usepackage{natbib}
\usepackage{hyperref}
\usepackage{lipsum}
\usepackage[ruled,vlined]{algorithm2e}

\begin{document}
%\title{Slow component resilience in timescale separated synchronized oscillators} 

\title{Resilience of the slow component in timescale separated synchronized oscillators} 
\author{Melvyn Tyloo}
\affiliation{Theoretical Division and Center for Nonlinear Studies (CNLS), Los Alamos National Laboratory, Los Alamos, NM 87545, USA}
\date{\today}

\begin{abstract}
Physiological networks are usually made of a large number of biological oscillators evolving on a multitude of different timescales. Phase oscillators are particularly useful in the modelling of the synchronization dynamics of such systems. If the coupling is strong enough compared to the heterogeneity of the internal parameters, synchronized states might emerge where phase oscillators start to behave coherently. Here, we focus on the case where synchronized oscillators are divided into a fast and a slow component so that the two subsets evolve on separated timescales. We assess the resilience of the slow component by reducing the dynamics of the fast one and evaluating the variance of the phase deviations when the oscillators in the two components are subject to noise with possibly distinct correlation times. From the general expression for the variance, we consider specific network structures and show how the noise transmission between the fast and slow components is affected. Interestingly, we find that oscillators that are among the most robust when there is only a single timescale, might become the most vulnerable when the system separates into two timescales. We also find that layered networks seem to be insensitive to timescale separations when the noise has homogeneous correlation time.
%%% Leave the Abstract empty if your article does not require one, please see the Summary Table for full details.

\end{abstract}
\maketitle

\section{Introduction}
The synchronization dynamics of coupled phase oscillators finds numerous applications ranging from Josephson junctions and electrical power grids to physiological networks~\cite{Wie98,Pik03,Ace05,strogatz2014nonlineardynamics,stiefel2016neurons}. The collective behavior displayed by these system is made possible by the interplay between the internal parameters of the individual dynamical units and the interaction coupling their degrees of freedom~\cite{winfree1967biological,Kur75,Kur84b,Str00}. Due to the nonlinear nature of the coupling together with the complex network topology of the interaction, multiple synchronized states might exist for the same parameters and might be visited by the system due to perturbations or noise~\cite{Kra40,Dyk90,DeV12,Rod16}. Importantly, synchronization is not always a desirable feature. For example, in electrical power grids, a synchronous operational state ensures the good functioning and distribution of power~\cite{Mac08,blaabjerg2006overview,Dor13}. The answer is less binary in physiological systems. Indeed, synchronization is of primal importance for some cognitive processes in the brain ensuring an adequate level of communication between neuronal groups~\cite{fries2005mechanism}. Also, synchronized dynamics emerge in healthy neuronal systems during sleep~\cite{gonzalez2023sleep}. Therefore, a lack of synchronization might result in some impairment of physiological system. However, direct connections have been drawn between the excess of synchronization in some neuronal groups and brain diseases~\cite{uhlhaas2006neural,popovych2014control}. Therefore, the synchronization dynamics as well as its resilience to external perturbations are topics of primal importance in order to better understand the interplay between synchronized groups of dynamical units.\\
Synchronization typically occurs in the phases or frequencies, where in the latter case all the oscillators are in a phase-locked state with time-independent phase differences, while in the former case all phases are the same. Perturbations of these synchronized states can take a great variety of forms such as external input signals injected into some internal parameters or noisy environments~\cite{DeV12,hindes2015driven,Scha17,Hin18,ronellenfitsch2018optimal,Tyl18a,halekotte2020minimal,tyloo2022layered}, interruption of the interaction between some oscillators due to local failures~\cite{Sol17,DELABAYS2022270}, alteration of the dynamics of some units~\cite{8320800,Tyloo_2023}. Here, we are interested in networks of phase oscillators in a phase-locked state where, due to damage to a subset of oscillators or simply because of their intrinsic characteristics, two separate timescales of the dynamics exist so that the system is divided into a fast and a slow component. This kind of timescale separation might occur for example in the human physiological system thanks to the wide range of timescales reported~\cite{gao2020neuronal}. In such a scenario, the fast oscillators adapt to any input signal quickly compared to the ones in the slow component. Therefore, the input signals into the oscillators belonging to the fast component are transmitted differently to those in the slow component compared to the case where all oscillators evolve on the same timescale. Such timescale separation in systems of coupled phase oscillators have been used in the modelling of power systems~\cite{kokotovic1980singular} and synchronization dynamics of Kuramoto oscillators with attractive and repulsive couplings~\cite{kirillov2020role}. As a paradigmatic model to investigate synchronization, we use Kuramoto oscillators, but the framework presented here applies more generally to coupled dynamical systems evolving close to a stable fixed point. We consider time-correlated noisy inputs as in many relevant situations, dynamical systems are constantly pushed away from their synchronized fixed point by ambient noisy conditions~\cite{Kam76}. The resilience of the system to such perturbations can be assessed in various ways. One can estimate the size of the basin of attraction~\cite{Wil06,Men13}, or evaluate the amplitude of small fluctuations or the escape rate of large fluctuations~\cite{Dyk90,DeV12,halekotte2020minimal,Scha17,Hin18,tyloo2022faster,hindes2023stability}. In this manuscript, we assess the resilience of the slow component in the small fluctuation regime by quantifying the phase deviations from the synchronized state. This is important as it clarifies which features of the dynamical system make it robust to noise when coupled oscillators evolve over multiple timescales. Within the assumption of small fluctuation, we investigate the linear response of the system around a stable fixed point. We first account for the timescale separation applying Mori-Zwanzig formalism~\cite{mori1965transport,zwanzig1973nonlinear} to the slow and fast components. This leads to a reduced dynamics of the oscillators which is equivalent to a Kron reduction of the Jacobian matrix~\cite{Kro39,dorfler2012kron}. The latter elucidate how the inputs in the fast component are transmitted to the slow one. Then, solving the linear system, we calculate the variance of the phase deviations in the slow component when time-correlated noises with distinct typical correlation times are applied in each component. We show how the amplitude of the excursion essentially depends on the characteristics of the noise, as well as the system properties through the spectrum of its reduced Jacobian. In some specific settings, we are able to further predict the transmission of the noise from the fast to the slow component based on the properties of the oscillators in the fast component as well as the inter-component coupling structure. In particular, we find that some oscillators having smaller variance when there is no timescale separation, might become the ones with larger variance when there is a timescale separation, if there a well connected to the fast component. Also, when the slow and fast components are defined on a layered network, the variance is mostly insensitive to the timescale separation when the noise correlation time is homogeneous.

In Sec.~\ref{sec2}, we introduce the model of Kuramoto oscillators with timescale separation and apply Mori-Zwanzig formalism to obtain a reduced dynamics for the slow component. In the same section, we then calculate the variance of the degrees of freedom of the oscillators in the slow component subject to time-correlated noise. In Sec.~\ref{sec3}, we numerically confirm and illustrate the theory on various network structures. The conclusions are given in Sec.~{\ref{sec4}}

\section{Timescale separation}\label{sec2}
\subsection{Networks of phase oscillators}
We are interested in the situation where, due to an external perturbation or change in the environment, the intrinsic timescales of the individual oscillators separate into a fast and a slow subsystem. We consider a set of $N$ oscillators each with a compact phase degree of freedom $\theta_i\in(-\pi,\pi]$ whose time-evolution is governed by the set of coupled differential equations~\cite{Kur75},
\begin{align}\label{eq1}
\begin{split}
    d_i\,\dot{\theta}_i = \omega_i - \sum_{j=1}^N b_{ij}\,\sin(\theta_i-\theta_j) + \eta_i\,,
\end{split}
\end{align}
for $i=1,...\,N$\,. The natural frequency of the $i$th oscillator is denoted $\omega_i$\,, the structure of the coupling network is given by elements $b_{ij}$ of the adjacency matrix~\cite{New18book}. Ambient noise is modelled at the $i$th oscillator by $\eta_i$ and is taken as a time-correlated noise, uncorrelated in space, i.e. $\langle \eta_i(t)\eta_j(t') \rangle = \eta_{0,i}^2\delta_{ij}\exp[-|t-t'|/\tau_i]$\,, where $\tau_i$ is the typical correlation time of the noise at the $i$th oscillator. The non-negative parameters $d_i$'s define the individual timescale of each oscillator. Removing the noise term, Eq.~(\ref{eq1}) may have multiple stable fixed points of the dynamics which essentially depend on the coupling topology and strength as well as the distribution of natural frequencies. Below, we assume that such a stable fixed point $\{\theta_i^{*}\}$ exists and that the noise term is small enough such that the dynamics remains inside the initial basin of attraction. In the present scenario, we assume that we have two sets of oscillators that we denote $\mathcal{F}$ and $\mathcal{S}$\,, respectively with $N_{\mathcal{F}}$ and $N_{\mathcal{S}}$ oscillators, such that 
\begin{eqnarray}
d_i=\begin{cases}
\overline{d} & i\in\mathcal{F}\\
\underline{d} & i\in\mathcal{S}
\end{cases}
\end{eqnarray}
with $\overline{d}\ll \underline{d}$\,. The latter means that oscillators belonging to $\mathcal{S}$ have a much slower intrinsic timescale than those belonging to $\mathcal{F}$\,. In the following, we focus on the dynamics of the oscillators in the slow component. Within the assumption of timescale separation, one can rewrite Eq.~(\ref{eq1}) as,
\begin{align}\label{eq2}
\begin{split}
    \,\dot{\theta}_i &= \omega_i - \sum_{j=1}^N b_{ij}\,\sin(\theta_i-\theta_j) + \eta_i\,, \,i\in\mathcal{S}\\
   \epsilon\,\dot{\theta}_i &= \omega_i - \sum_{j=1}^N b_{ij}\,\sin(\theta_i-\theta_j) + \eta_i\,, \,i\in\mathcal{F}\,,
\end{split}
\end{align}
where we defined $\underline{d}/\overline{d}=\epsilon^{-1}$ and, without loss of generality, set $\underline{d}$ to unity. In the limit $\epsilon\rightarrow 0$\,, the oscillators within $\mathcal{F}$ instantaneously adapt their phases. In the next section, we consider the dynamical system Eqs.~(\ref{eq2}) in the vicinity of a stable fixed point and perform a singular perturbation analysis using Mori-Zwanzig formalism.
\subsection{Near-equilibrium and reduced dynamics}
Even though we consider Kuramoto oscillators, the following approach applies in general to coupled dynamical systems that have a stable fixed point around which they evolve and where linearization is valid. To analyze the resilience of the slow component, we consider the dynamics of the system close to a fixed point $\{\theta_i^{*}\}$\,. In particular, we are interested in the time-evolution of the phase deviations $x_i(t)=\theta_i(t)-\theta_i^{*}$ for $i\in\mathcal{S}$ and $y_i(t)=\theta_i(t)-\theta_i^{*}$ for $i\in\mathcal{F}$ whose dynamics at the first order reads,
\begin{align}\label{eq4}
\begin{bmatrix}
\dot{\bf x}\\
\epsilon\,\dot{\bf y}
\end{bmatrix}=
\begin{bmatrix}
{\bf J}_{\mathcal{S}\mathcal{S}} & {\bf J}_{\mathcal{S}\mathcal{F}}\\
{\bf J}_{\mathcal{F}\mathcal{S}} & {\bf J}_{\mathcal{F}\mathcal{F}}
\end{bmatrix}
\begin{bmatrix}
{\bf x}\\
{\bf y}
\end{bmatrix}+
\begin{bmatrix}
{\bm \eta}_\mathcal{S}\\
{\bm \eta}_\mathcal{F}
\end{bmatrix}
\end{align}
where we defined the matrix
\begin{eqnarray}\label{eqjac}
J_{ij}=\begin{cases}
 b_{ij}\cos(\theta_i^*-\theta_j^*)& i\neq j\\
 -\sum_{k=1}^Nb_{ik}\cos(\theta_i^*-\theta_k^*)& i=j\,,
\end{cases}
\end{eqnarray}
which is the Jacobian of the system and is a Laplacian matrix when phase differences are between $-\frac{\pi}{2}$ and $\frac{\pi}{2}$\,. Using Mori-Zwanzig formalism~\cite{mori1965transport,zwanzig1973nonlinear} with $\bf x$ and $\bf y$ being respectively the resolved and unresolved variables (see \cite{caravelli2023combinatorics} for an introduction), one can express the first row of Eq.~(\ref{eq4}) as,
\begin{align}\label{eqmz}
    \dot{x}_i &= \sum_{j=1}^{N_{\mathcal{S}}}{J_{\mathcal{S}\mathcal{S}}}_{ij}\,x_j + {\eta_{\mathcal{S}}}_i + \sum_{\alpha=1}^{N_{\mathcal{F}}}\int_0^t \epsilon^{-1}e^{\nu_\alpha\epsilon^{-1}(t-t')} \\
    &\times\sum_{k,m=1}^{N_{\mathcal{F}}}\sum_{l=1}^{N_{\mathcal{S}}}{J_{\mathcal{FS}}}_{kl}x_l(t')w_{\alpha,k}{J_{\mathcal{SF}}}_{im}w_{\alpha,m}\, {\rm d}t' \nonumber \\ \nonumber
    + &\sum_{\alpha=1}^{N_{\mathcal{F}}}\int_0^t \epsilon^{-1}e^{\nu_\alpha\epsilon^{-1}(t-t')}\sum_{k,m=1}^{N_{\mathcal{F}}}{\eta_{\mathcal{F}}}_k(t')w_{\alpha,k}{J_{\mathcal{SF}}}_{im}w_{\alpha,m}\, {\rm d}t'\,,
\end{align}
where we denoted ${\bf w}_{\alpha}$ the eigenvectors of ${\bf J}_{\mathcal{F}\mathcal{F}}$ with corresponding eigenvalues $\nu_\alpha<0$\,. In Eq.~(\ref{eqmz})\,, the first two terms on the right-hand side are Markovian while the other ones have memory. 
We are interested in the time-evolution of the slow components ${\bf x}$ when there is a timescale separation, i.e. $\epsilon\rightarrow 0$\,. Taking the latter limit in Eq.~(\ref{eqmz}) yields in a matrix form,
\begin{align}\label{eq5}
    \dot{\bf x}&={\bf J}_{\mathcal{S}\mathcal{S}}\,{\bf x} - {\bf J}_{\mathcal{S}\mathcal{F}}\, {\bf J}^{-1}_{\mathcal{F}\mathcal{F}}\,{\bf J}_{\mathcal{F}\mathcal{S}}\,{\bf x} +  {\bm \eta}_{\mathcal{S}}- {\bf J}_{\mathcal{S}\mathcal{F}}\, {\bf J}^{-1}_{\mathcal{F}\mathcal{F}}\,{\bm \eta}_{\mathcal{F}}\\
    &={\bf J}_{\rm red}\,{\bf x} + {\bm \xi}  \,,\nonumber
\end{align}
where in the second line we defined the reduced Jacobian ${\bf J}_{\rm red}={\bf J}_{\mathcal{S}\mathcal{S}} - {\bf J}_{\mathcal{S}\mathcal{F}}\, {\bf J}^{-1}_{\mathcal{F}\mathcal{F}}\,{\bf J}_{\mathcal{F}\mathcal{S}}$\,, and denoted the noise term as ${\bm \xi}={\bm \eta}_{\mathcal{S}}- {\bf J}_{\mathcal{S}\mathcal{F}}\, {\bf J}^{-1}_{\mathcal{F}\mathcal{F}}\,{\bm \eta}_{\mathcal{F}}$\,. It is interesting to note that the reduced dynamics given by Eq.~(\ref{eq5}) can be obtained by a Kron reduction~\cite{Kro39,dorfler2012kron} of the fast component of the system (see \cite{tyloo2023forced} for an example).
The dynamics of the slow component is then governed by Eq.~(\ref{eq5}) where the effective noise at the $i$th oscillator is a combination of the noise at the $i$th oscillator combined with a superposition of the noise inputs at oscillators belonging to the fast component. For undirected coupling as we consider in the following, one has that ${\bf J}_{\mathcal{F}\mathcal{S}}={\bf J}_{\mathcal{S}\mathcal{F}}^\top$\,. The linear system Eq.~(\ref{eq5}) can be solved by expanding over the eigenmodes of ${\bf J}_{\rm red}$ denoted ${\bf u}_{\alpha}$\,, with corresponding eigenvalues $\lambda_\alpha$\,, $\alpha=1,...N_{\mathcal{S}}=|\mathcal{S}|$\,. As ${\bf J}_{\rm red}$ is also the negative of a Laplacian matrix, one has that $0>\lambda_1\ge ... \ge \lambda_{N_{\mathcal{S}}}$ with $u_{1,i}=1/\sqrt{N_{\mathcal{S}}}$\,. The general solution to Eq.~(\ref{eq5}) is given by,
\begin{align}\label{eqsol}
x_i(t) = \sum_{\alpha=2}^{N_{\mathcal{S}}} \int_{0}^{t}\exp[\lambda_\alpha(t-t')]{\bf u}_{\alpha}\cdot{\bm \xi}(t')\,{\rm d}t'u_{\alpha,i}\,,
\end{align}
for $ i=2,...N_{\mathcal{S}}$ and where we omitted the first mode in the sum as any perturbation along it does not modify the system\,. This expression can be used directly to calculate the moments of the phase deviations.

\begin{figure}
\centering
\includegraphics[scale=1.9]{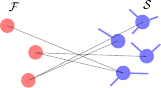}
\caption{Connectivity where oscillators in the fast component are disconnected. The fast oscillators are shown in red while the slow oscillators are shown in blue.}\label{fig11}
\end{figure}
\subsection{Fluctuations from the synchronized state}
Various characteristics of the response can be used to determine the resilience of the coupled oscillators. When subject to stochastic inputs, a natural choice is to evaluate the magnitude of the deviations from the synchronized fixed point by calculating the variance of the phase deviations. The variance of the phase deviations in the slow component is calculated from Eq.~(\ref{eqsol}) and reads in the long time limit,
\begin{align}\label{eq6}
\langle x_i^2 \rangle &=\eta_{0,\mathcal{S}}^2 \sum_{\alpha=2}^{N_{\mathcal {S}}}\frac{u_{\alpha,i}^2}{\lambda_\alpha(\lambda_\alpha-\tau_\mathcal{S}^{-1})}\\
+ \eta_{0,\mathcal{F}}^2&\sum_{\alpha,\beta=2}^{N_{\mathcal {S}}}\frac{(\lambda_\alpha + \lambda_\beta - 2\tau_\mathcal{F}^{-1})\Gamma_{\alpha\beta}\,u_{\alpha,i}u_{\beta,i}}{(\lambda_\alpha + \lambda_\beta)(\tau_\mathcal{F}^{-1}-\lambda_\alpha)(\tau_\mathcal{F}^{-1}-\lambda_\beta)}\,,\nonumber
\end{align}
with  the scalar $\Gamma_{\alpha\beta}={\bf u}^\top_{\alpha}{\bf J}_{\mathcal{S}\mathcal{F}}{\bf J}_{\mathcal{F}\mathcal{F}}^{-2} {\bf J}_{\mathcal{F}\mathcal{S}}{\bf u}_{\beta}$\,. In Eq.~(\ref{eq6}), we set the standard deviation of the ambient noise in the slow and fast components respectively, to $=\eta_{0,\mathcal{S}}$ and $\eta_{0,\mathcal{F}}$\,. We also set distinct homogeneous correlation times for the noise in each component as $\tau_i=\tau_\mathcal{S}$ for $i\in \mathcal{S}$ and $\tau_i=\tau_\mathcal{F}$ for $i\in \mathcal{F}$ \,.
While the contribution to the variance from the additive noise in the slow component is essentially given by the position of the oscillators on the slowest eigenmodes, the effect of the noise coming from the fast component involves combinations of eigenmodes. The precise combination depends on the effective reduced dynamics through $\Gamma_{\alpha\beta}$\,.
The shortest timescale in the system is set by the oscillators belonging to $\mathcal{F}$\,. However, by tuning the correlation time of the noise $\tau=\tau_{\mathcal{F}}=\tau_{\mathcal{S}}$\,, one can investigate the regimes where the $\lambda_{N_\mathcal{S}}\tau\ll 1$ and $\lambda_2\tau\gg 1$\,. Indeed, in the limit where the noise correlation time is shorter than the timescales of the slow component and the same in both components, the variance becomes,
\begin{align}\label{eq7}
\langle x_i^2 \rangle &=\eta_{0,\mathcal{S}}^2\,\tau \sum_{\alpha=2}^{N_{\mathcal {S}}}\frac{u_{\alpha,i}^2}{(-\lambda_\alpha)}+ \eta_{0,\mathcal{F}}^2\,\tau\sum_{\alpha,\beta=2}^{N_{\mathcal {S}}}\frac{2\,\Gamma_{\alpha\beta}\,u_{\alpha,i}u_{\beta,i}}{(-\lambda_\alpha - \lambda_\beta)}\,.
\end{align}
In the other limit where the noise correlation time is the longest timescale, one has,
\begin{align}\label{eq8}
\langle x_i^2 \rangle &=\eta_{0,\mathcal{S}}^2\, \sum_{\alpha=2}^{N_{\mathcal {S}}}\frac{u_{\alpha,i}^2}{\lambda_\alpha^2} + \eta_{0,\mathcal{F}}^2\,\sum_{\alpha,\beta=2}^{N_{\mathcal {S}}}\frac{\Gamma_{\alpha\beta}\,u_{\alpha,i}u_{\beta,i}}{\lambda_\alpha \lambda_\beta}\,.
\end{align}
Comparing the two limiting cases Eqs.~(\ref{eq7}) and (\ref{eq8})\,, one remarks that in both variances, a significant contribution might come from the slowest eigenmodes. Note also that Eq.~(\ref{eq6}) is more generally valid in the case where $\tau_{\mathcal{F}}$ and $\tau_{\mathcal{S}}$ are different.

In the Appendix~\ref{app1}\,, we give the variance of the phases when there is no timescale separation.

To obtain more insights into the contribution from the fast component, let us consider specific situations in the strong coupling limit.

\subsection{Strong coupling limit}\label{secStr}
In the strong coupling limit, one has $|\theta_i^*-\theta_j^*|\ll 1$ $\forall i,j$\,, so that one can approximate the Jacobian Eq.~(\ref{eqjac}) as,
\begin{eqnarray}\label{eqjac1}
J_{ij}=\begin{cases}
 b_{ij}& i\neq j\\
 -\sum_{k=1}^N b_{ik}& i=j\,.
\end{cases}
\end{eqnarray}
Within this coupling limit and some other assumptions that are specified below, one can further consider network structures that give more insights about Eq.~(\ref{eq6})\,.

\begin{figure}
\centering
\includegraphics[scale=1.9]{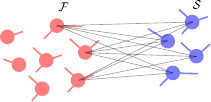}
\caption{Connectivity where oscillators in the fast component with inter-component connections are connected to all the oscillators in the slow component. The fast oscillators are shown in red while the slow oscillators are shown in blue.}\label{fig1}
\end{figure}
\begin{figure}
\centering
\includegraphics[scale=1.9]{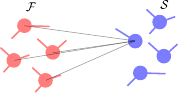}
\caption{Connectivity where a single oscillator in the slow component is connected to all the oscillators in the fast one. The fast oscillators are shown in red while the slow oscillators are shown in blue.}\label{fig12}
\end{figure}
\subsubsection{Disconnected oscillators in the fast component}
In the simple scenario where only a single oscillator $l$ belongs to the fast component while all the others are in the slow one, ${\bf J}_{\mathcal{F}\mathcal{F}}$ is a scalar such that ${J}_{\mathcal{F}\mathcal{F}}^{-2}=k_{l}^{-2}$ is the inverse of the squared weighted degree of the fast oscillator indexed by $l$\,. Therefore, one has $\Gamma_{\alpha\beta}=\left(\sum_{j\in\mathcal{N}(l)}u_{\alpha,j}b_{lj}\right)\left(\sum_{k\in\mathcal{N}(l)}u_{\beta,k}b_{lk}\right)/k_{l}^2$\,, where $\mathcal{N}(l)$ is the set of oscillators in the slow component connected to the fast oscillator $l$\,. The contribution from the second term in Eq.~(\ref{eq6}) therefore crucially depends on the location of the oscillators on the slowest eigenmodes of the reduced Jacobian. This situation easily generalizes to the case of multiple oscillators in the fast component that are not connected as shown in Fig.~\ref{fig11}\,. One then has $\Gamma_{\alpha\beta}=\sum_{l\in\mathcal{F}}\left(\sum_{j\in\mathcal{N}(l)}u_{\alpha,j}b_{lj}\right)\left(\sum_{k\in\mathcal{N}(l)}u_{\beta,k}b_{lk}\right)/k_{l}^2$\,, where we took the sum over all the oscillators in the fast component.

\subsubsection{All-to-all coupling from fast to slow component}\label{fts}
When the oscillators in the fast components that have inter-component connections are connected to all the oscillators in the slow component with homogeneous coupling, i.e. $b_{lj}=b_l>0$ $\forall j\in \mathcal{N}(l)=\mathcal{S}$\,, with $l\in\mathcal{F}$ (see Fig.~{\ref{fig1}), the second term in Eq.~(\ref{eq6}) vanishes. Indeed, in such a situation, the columns of the matrix ${\bf J}_{\mathcal{S}\mathcal{F}}={\bf J}_{\mathcal{F}\mathcal{S}}^\top$ corresponding to the inter-component coupling are full of ones and as ${\bf u}_1\cdot {\bf u}_\alpha=\delta_{1\alpha}$ by orthogonality of the eigenmodes, one has that $\Gamma_{\alpha\beta}=0$ for all $\alpha,\beta=1,...,N_{\mathcal{S}}$\,. Intuitively, if the signal from one oscillator in the fast component is transmitted with the same strength to all the oscillators in the slow one, then it will result in an overall phase shift without affecting the fixed point. However, if the coupling is not homogeneous between the slow and fast components, the variance will be different from zero.

\subsubsection{All-to-all from slow to fast component}\label{fts2}
In the opposite case where only a single oscillator in the slow component is homogeneously connected to all the oscillators in the fast one (see Fig.~\ref{fig12}), the intra-component structure of the coupling within the fast component does not influence the propagation of the noise. Indeed, if $b_{lj}=b_l>0$ $\forall j\in \mathcal{M}(l)=\mathcal{F}$ where here $\mathcal{M}(l)$ is the set of oscillators in the fast component connected to the $l$th oscillator in the slow component, one has $\Gamma_{\alpha\beta}= b_{l}^2u_{\alpha,l}u_{\beta,l}N_{\mathcal{F}}$\,. Therefore, only the size of the fast component and not the intra-component network structure of the oscillators influence the variance in the slow one. This can be generalized to the case where multiple oscillators in slow component are connected to all the ones in the fast component. One then has, $\Gamma_{\alpha\beta}= \sum_{l\in\mathcal{S}}b_{l}^2u_{\alpha,l}u_{\beta,l}m^{-2}$ with $m$ the number of oscillators in the slow component that have all-to-all connections to the fast component. Here again, the structure of the coupling within the fast component does not influence the variance.

\subsubsection{Layered networks}\label{secl}
An interesting case arises when oscillators are connected on layered networks so that the fast component is on one layer, the slow one on another layer, and the two layers are connected together. In the specific scenario where each fast oscillator is connected to a single distinct oscillator in the slow component and the number of units in the layers is the same, one has that ${\bf J}_{\mathcal{S}\mathcal{F}}$ is a diagonal matrix (up to a permutation of the oscillators indices). This is illustrated in Fig.~\ref{figlayer}\,. If one further assumes that the inter-layer coupling is homogeneous, i.e.  ${\bf J}_{\mathcal{S}\mathcal{F}}=\tilde{b}\,\mathbb{I}$ is a multiple of the identity matrix, one has that the eigenbases of ${\bf J}_{\rm red}$ and ${\bf J}_{\mathcal{S}\mathcal{S}}={\bf J}_{\mathcal{F}\mathcal{F}}$ satisfy the following relations,
\begin{align}
{\bf J}_{\rm red}{\bf u}_{\alpha}=\lambda_\alpha {\bf u}_{\alpha} &= ({\bf J}_{\mathcal{S}\mathcal{S}} - \tilde{b}^2{{\bf J}_{\mathcal{S}\mathcal{S}}}^{-1}){\bf u}_\alpha\,,\\
({\bf J}_{\mathcal{S}\mathcal{S}} - \tilde{b}^2{{\bf J}_{\mathcal{S}\mathcal{S}}}^{-1}){\bf v}_\beta &= (\mu_\beta - \tilde{b}^2\mu_\beta^{-1}){\bf v}_\beta = {\bf J}_{\rm red}{\bf v}_{\beta}\,,
\end{align}
with $\alpha,\beta=1,...,N_{\mathcal{S}}$ and where the eigenmodes of ${\bf J}_{\mathcal{S}\mathcal{S}}$ are denoted ${\bf v}_{\beta}$ with corresponding eigenvalue $\mu_\beta$\,. Noticing that $(\mu_\beta - \tilde{b}^2\,\mu_\beta^{-1})$ is a monotonically increasing function of $\mu_\beta$\,, one has a one-to-one correspondence between the eigenmodes of ${\bf J}_{\rm red}$ and ${\bf J}_{\mathcal{S}\mathcal{S}}$ such that ${\bf u}_\alpha={\bf v}_\alpha$ and $\lambda_\alpha=(\mu_\alpha - \tilde{b}^2\,\mu_\alpha^{-1})$ for  $\alpha=1,...,N_{\mathcal{S}}$\,. Given the specific structure of the coupling, one further has that $\mu_\beta = \gamma_\beta - \tilde{b}$ where $\gamma_\beta$ are the eigenvalues of the Jacobian in each of the layer when removing the inter-layer coupling.
%The smallest eigenvalue of ${\bf J}_{\rm red}$ in absolute value is therefore given by,
%\begin{align}
%\lambda_{2} = \min_{\beta}\left|\frac{\gamma_\beta(\gamma_\beta - 2\tilde{b})}{\gamma_\beta - \tilde{b}}\right|\,.
%\end{align}
%Interestingly, by selecting a repulsive inter-layer coupling $\tilde{b}<0$ approaching $|\gamma_2|/2$ from below, one may have $\lambda_2$ that becomes closer and closer to zero, meaning that variance is more and more significant. In order for the system to remain stable, one needs $|\tilde{b}|<|\gamma_2|/2$\,. 
Assuming that the noise amplitudes as well as the correlation times are the same in both components, one can rewrite Eq.~(\ref{eq6}) as,
\begin{align}\label{eq11}
\langle x_i^2 \rangle &=\eta_{0}^2 \sum_{\alpha=2}^{N_{\mathcal {S}}}\frac{v_{\alpha,i}^2(\mu_\alpha^2+\tilde{b}^2)}{(\mu_\alpha^2-\tilde{b}^2)(\mu_\alpha^2 - \tilde{b}^2-\mu_\alpha\tau^{-1})}\,.
\end{align}
One can also calculate the variance when there is no timescale separation by remarking that the eigenmodes of the full Jacobian Eq.~(\ref{eqjac}) are given by $[{\bf v}_\alpha, \pm {\bf v}_\alpha]^\top/\sqrt{2}$ with corresponding eigenvalues $(\mu_\alpha \pm \tilde{b})$ for $\alpha=1,...,(N_{\mathcal{S}}+N_{\mathcal{F}})$\,. Using Eq.~(\ref{eqapp})\,, the variance then reads,
\begin{align}\label{eq12}
\langle x_i^2 \rangle &=\eta_{0}^2 \sum_{\alpha=2}^{N_{\mathcal {S}}}\frac{v_{\alpha,i}^2(-\mu_\alpha\tau^{-1}+\mu_\alpha^2+\tilde{b}^2)}{(\mu_\alpha^2-\tilde{b}^2)(\tau^{-2}-2\mu_\alpha\tau^{-1} - \tilde{b}^2 +\mu_\alpha^2)} \\
&+ \frac{\eta_0^2(N_{\mathcal{S}}+N_{\mathcal{F}})^{-1}}{2\tilde{b}(\tau^{-1}+2\tilde{b})} \nonumber\,.
\end{align}
While Eqs.~(\ref{eq11}), (\ref{eq12}) are different in general, when the correlation time of the noise becomes short, the two variances only differ by a constant given by the second term in the right-hand side of Eq.~(\ref{eq12})\,.

\begin{figure}
\centering
\includegraphics[scale=1.9]{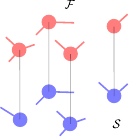}
\caption{Connectivity in layers where the fast oscillators are shown in red in the top layer while the slow oscillators are shown in blue in the bottom layer. Each oscillator in one of the layer can be connected only to a single oscillator in the other layer.}\label{figlayer}
\end{figure}

%\begin{align}\label{eq9}
%\langle \delta\theta_i^2 \rangle &=\eta_{0,\mathcal{S}}^2 \sum_{\alpha=2}^{N_{\mathcal {S}}}\frac{v_{\beta,i}^2}{(\mu_\beta - \tilde{b}^2\mu_\beta^{-1})[(\mu_\beta - \tilde{b}^2\mu_\beta^{-1})-\tau^{-1}]}\\
%- \eta_{0,\mathcal{F}}^2&\sum_{\beta=2}^{N_{\mathcal {S}}}\frac{v_{\beta,i}^2}{\mu_\beta^2(\mu_\beta - \tilde{b}\mu_{\beta}^{-1})(\tau^{-1}-\mu_\beta - \tilde{b}^2\,\mu_\beta^{-1})}\,,\nonumber
%\end{align}

In the following section, we illustrate and confirm numerically the results discussed so far.
\begin{figure}
\centering
\includegraphics[scale=0.55]{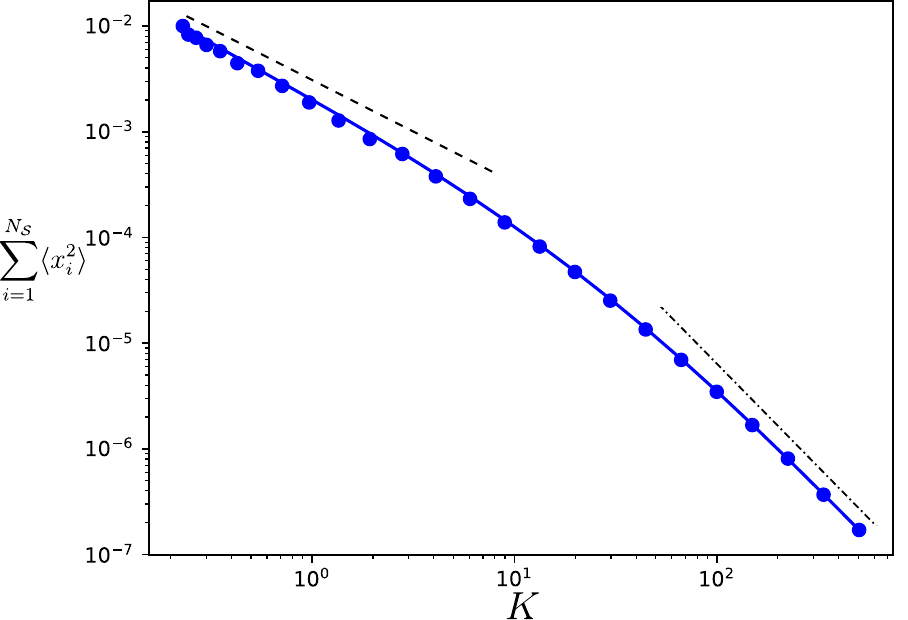}
\caption{\small Sum of the variance of the phase deviations at every oscillator in the slow component as a function of the coupling strength $K$\,. The variance is obtained by time-evolving Eq.~(\ref{eq1}) for a Watts-Strogatz network of 50 oscillators with heterogeneous natural frequencies (${\rm var}[\omega]=0.01$) where the ratio of the damping parameters in the slow and fast component is $\overline{d}/\underline{d}=0.1$\,. The correlation times of the noise are fixed to $\tau_{\mathcal{S}}/\underline{d}=\tau_{\mathcal{F}}/\underline{d}=0.05$\,. Each dot is obtained by time-averaging the variance over a single simulation of the dynamics while the solid curve gives the theory that is the sum of the variances in Eq.~(\ref{eq6})\,. The noise amplitudes are $\eta_{0,\mathcal{S}}/\underline{d}=\eta_{0,\mathcal{F}}/\underline{d}=0.05$\,. The black dashed and dotted-dashed lines give respectively, $\sim K^{-1}$ and $\sim K^{-2}$\,.}\label{thnum}
\end{figure}
\section{Numerical Results}\label{sec3}
We first confirm numerically Eq.~(\ref{eq6}) in Fig.~\ref{thnum} where we consider a Watts-Strogatz network~\cite{New18book} with $m=4$ initial nearest neighbors and a rewiring probability $p_{\rm rewiring}=0.1$\,, of size $N_\mathcal{S}+N_\mathcal{F}=50$\,, with respectively $N_{\mathcal{S}}=35$ and $N_{\mathcal{F}}=15$ oscillators in each component. The damping parameters are taken homogeneous in each component and satisfy $\overline{d}/\underline{d}=0.1$ in order to effectively simulate a timescale separated system\,. We fix the correlation times of the noise $\tau_{\mathcal{S}}/\underline{d}=\tau_{\mathcal{F}}/\underline{d}=0.05$ and tune the coupling weights of the network as $b_{ij}\rightarrow K\,b_{ij}$ $\forall i,j=1,...,50$\, in order to visit the different regimes. As predicted by the theory Eq.~(\ref{eq6}) and its two limiting cases Eqs.~(\ref{eq7}), (\ref{eq8})\,, when the noise correlation time is shorter than the system's timescales, one has that $\sum_{i\in\mathcal{S}}\langle x_i^2 \rangle\sim 1/K$ while in the case where the noise is much slower than the system's timescales one has $\sum_{i\in\mathcal{S}}\langle x_i^2 \rangle\sim 1/K^2$\,. Moreover, the theory given by the solid line matches the numerical results given by the dots.
\begin{figure}
\centering
\includegraphics[scale=0.52]{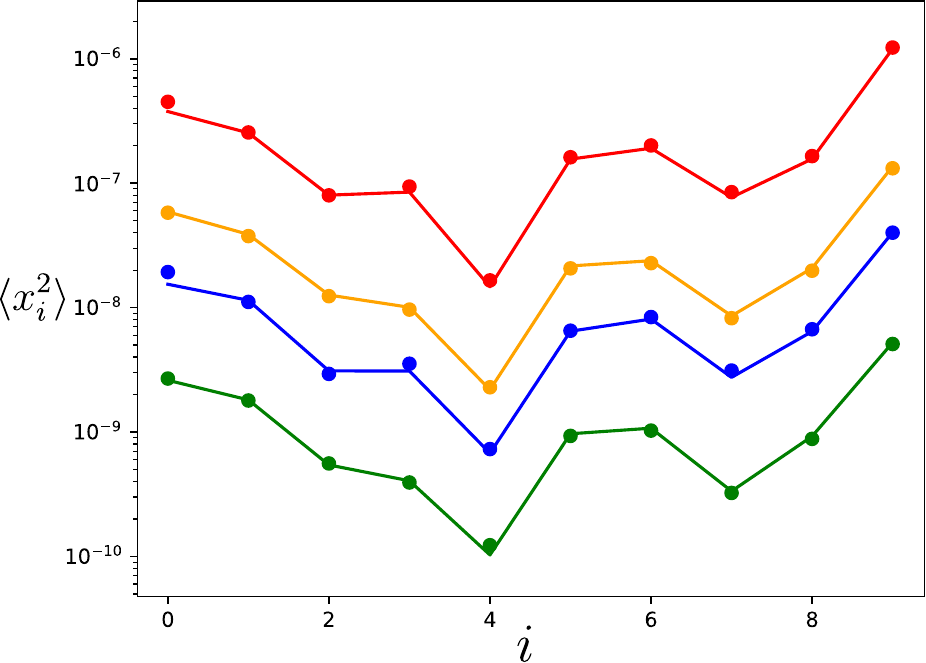}
\caption{\small Variance of the phase deviations at every oscillator in the slow component when noise is only applied on the fast component. The variance is obtained by time-evolving Eq.~(\ref{eq1}) for a modified Erd\H{o}s-R\'enyi network of 13 oscillators where 3 oscillators in the fast component are connected to all the oscillators in the slow component. The ratio of the damping parameters in the slow and fast component is $\overline{d}/\underline{d}=0.1$\,. The correlation time of the noise in the fast component is fixed to $\tau_{\mathcal{F}}/\underline{d}=0.05$ for the orange and green data points and $\tau_{\mathcal{F}}/\underline{d}=10$ for the red and blue ones. The natural frequencies are such that ${\rm var}[\omega]=0.01$ for the green and blue points and ${\rm var}[\omega]=0.05$ for the red and orange ones. Each dot is obtained by time-averaging the variance over a single simulation of the dynamics while the solid curve gives the theory Eq.~(\ref{eq6})\,. The noise amplitude is $\eta_{0,\mathcal{F}}/\underline{d}=0.1$\,.}\label{thnum2}
\end{figure}

\begin{figure}
\centering
\includegraphics[scale=0.55]{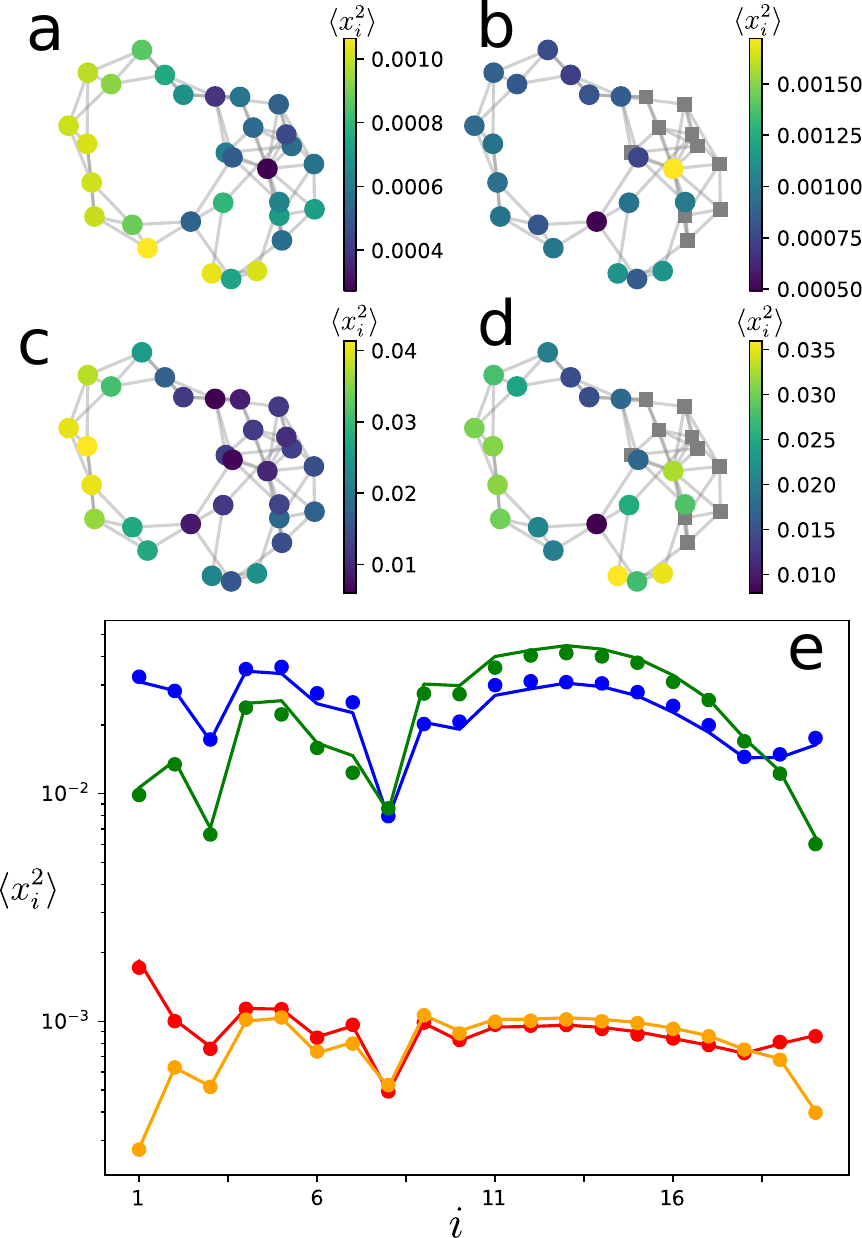}
\caption{\small Variance of the phase deviations at every oscillator when there no timescale separation [panels (a), (c) and orange and green data points in panel (e)] and when there is a timescale separation [panels (b), (d) and red and blue data points in panel (e)], for a modified Watts-Strogatz network of $N_\mathcal{S}+N_\mathcal{F}=30$ where the yellow node (with largest variance) in (b) is connected to all the oscillators in the fast component. The oscillators depicted by grey squares are in the fast component while all the others are in the slow component. The variance is numerically obtained by time-averaging over a single realization of the dynamics Eq.~(\ref{eq1})\,. (e) Comparison between the theory Eq.~(\ref{eq6}) given by the solid lines and the numerical simulations given by the dots when there is no timescale separation (in orange and green) and when there is a timescale separation (in red and blue). The ratio of the damping parameters in the slow and fast component is $\overline{d}/\underline{d}=0.02$\,. The correlation times of the noise are fixed to $\tau_{\mathcal{F}}/\underline{d}=\tau_{\mathcal{F}}/\underline{d}=0.05$ (orange and red) and $\tau_{\mathcal{F}}/\underline{d}=\tau_{\mathcal{F}}/\underline{d}=50$ (green and blue). The natural frequencies are such that ${\rm var}[\omega]=0.03$\,. The noise amplitude is $\eta_{0,\mathcal{F}}/\underline{d}=0.2$\,.}\label{thnum4}
\end{figure}
\begin{figure}
\centering
\includegraphics[scale=0.55]{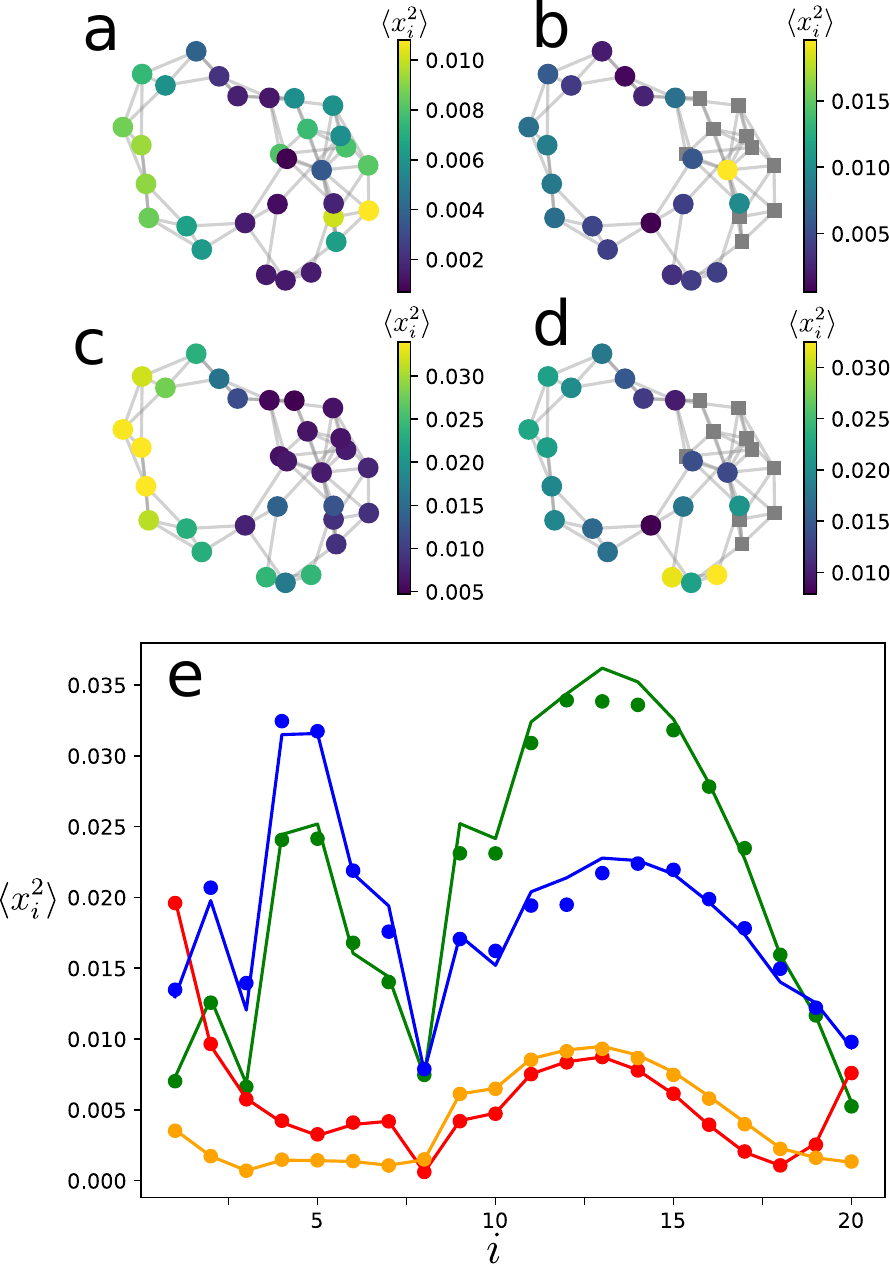}
\caption{\small Variance of the phase deviations at every oscillator when there no timescale separation [panels (a), (c) and orange and green data points in panel (e)] and when there is a timescale separation [panels (b), (d) and red and blue data points in panel (e)], for a modified Watts-Strogatz network of $N_\mathcal{S}+N_\mathcal{F}=30$ where the yellow node (with largest variance) in (b) is connected to all the oscillators in the fast component. The oscillators depicted by grey squares are in the fast component while all the others are in the slow component. The variance is numerically obtained by time-averaging over a single realization of the dynamics Eq.~(\ref{eq1})\,. (e) Comparison between the theory Eq.~(\ref{eq6}) given by the solid lines and the numerical simulations given by the dots when there is no timescale separation (in orange and green) and when there is a timescale separation (in red and blue). The ratio of the damping parameters in the slow and fast component is $\overline{d}/\underline{d}=0.02$\,. The correlation times of the noise are fixed to $\tau_{\mathcal{F}}/\underline{d}=0.05$\, $\tau_{\mathcal{F}}/\underline{d}=50$ (orange and red) and $\tau_{\mathcal{F}}/\underline{d}=50$\,, $\tau_{\mathcal{F}}/\underline{d}=0.05$ (green and blue). The natural frequencies are such that ${\rm var}[\omega]=0.03$\,. The noise amplitude is $\eta_{0,\mathcal{F}}/\underline{d}=0.2$\,.}\label{thnum5}
\end{figure}

\begin{figure}
\centering
\includegraphics[scale=0.55]{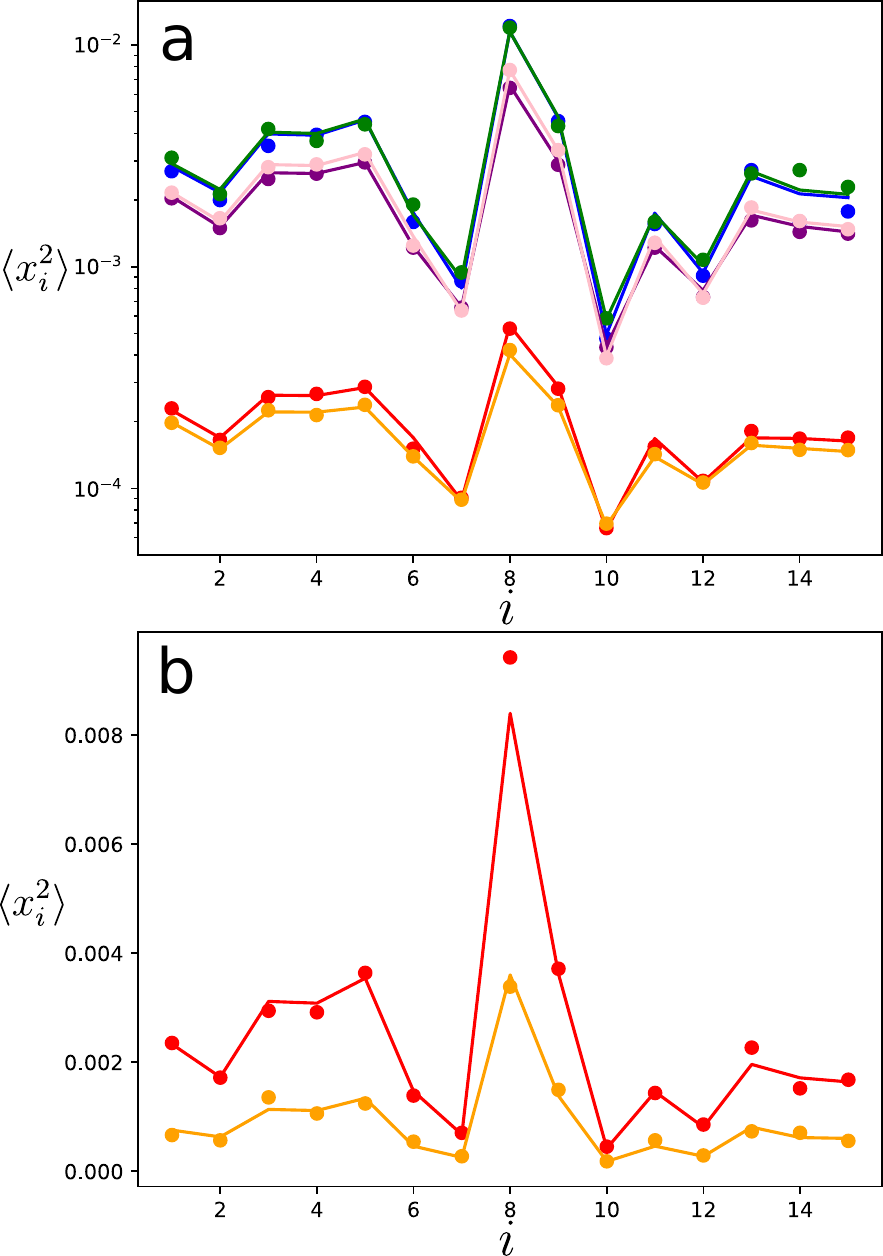}
\caption{\small Variance of the phase deviations at every oscillator in the slow component when noise is only applied on both the slow and the fast components. The variance is obtained by time-evolving Eq.~(\ref{eq1}) for a layered network made of two copies of an Erd\H{o}s-R\'enyi network of 15 oscillators. One layer is the fast component while the other is the slow one. Pairs of corresponding oscillators in the two layers have a single inter-component connection. The correlation time of the noise in both components is fixed to $\tau/\underline{d}=0.05$ for the red and orange data points, $\tau/\underline{d}=50$ for the green and blue ones, and $\tau/\underline{d}=2$ for the purple and pink ones. The natural frequencies are identical and vanishing for all oscillators. Each dot is obtained by time-averaging the variance over a single simulation of the dynamics while the solid curve gives the theory Eq.~(\ref{eq6}) when there is no timescale separation ($\overline{d}/\underline{d}=1$) for the orange, green and purple data points, and when there is a timescale separation ($\overline{d}/\underline{d}=0.05$) for the red, blue and pink ones. The noise amplitude is $\eta_{0,\mathcal{F}}/\underline{d}=0.1$\,.}\label{thnum3}
\end{figure}
Next, in Sec.~\ref{secStr}\,, we discuss particular examples in the strong coupling limit. We numerically investigate these examples also going beyond the strong coupling limit. 
\subsection{All-to-all coupling from fast to slow component}
We consider the situation described in Sec.~\ref{fts} where oscillators in the fast component that have some connection to the slow one are connected to all of them. The numerical results are shown on Fig.~\ref{thnum2} where this particular setting has been simulated for a modified Erd\H{o}s-R\'enyi network of $N_\mathcal{S}+N_\mathcal{F}=13$ oscillators with various correlation time of the noise and different heterogeneity levels in the natural frequencies. Here, only the fast component is subjected to noise, while the slow component is noiseless. One observes that increasing the heterogeneity in the natural frequencies of the oscillators induces larger variances for the phase deviations. Not shown on Fig.~\ref{thnum2} is the homogeneous case of oscillators with identical natural frequencies, for which the variance vanishes as predicted in Sec.~\ref{fts}\,. While the heterogeneity increases the noise transmission from the fast to the slow component, one observes that the amplitude of the deviations is still rather small in Fig.~\ref{thnum}\,. Besides ${\bf u}^\top_{\alpha}{\bf J}_{\mathcal{S}\mathcal{F}}$ being small, this is because the oscillators in the fast component with inter-component connections have relatively large degrees which directly reduce the noise transmission in $\Gamma_{\alpha\beta}={\bf u}^\top_{\alpha}{\bf J}_{\mathcal{S}\mathcal{F}}{\bf J}_{\mathcal{F}\mathcal{F}}^{-2} {\bf J}_{\mathcal{F}\mathcal{S}}{\bf u}_{\beta}$\,.

\subsection{All-to-all from slow to fast component}
In the other situation where some oscillators in the slow components are connected to a large fraction of the oscillators in the fast component, we showed in Sec.~\ref{fts2} that their variance is more important than oscillators with fewer or no connection to the fast component. This result is particularly interesting and intriguing, as in the regular situation where there is no timescale separation, oscillators that have a larger number of connections to the other elements typically have a smaller variance~\cite{Tyl19}. Indeed, this is first illustrated in Fig.~\ref{thnum4} where the variance of each oscillator is given by the color map when there is no timescale separation in panels (a) and (c), and when there is a timescale separation in panels (b) and (d) within the synchronized oscillators. In both correlation time limits (with $\tau_{\mathcal{S}}=\tau_{\mathcal{F}}$), one observes that oscillators belonging to $\mathcal{S}$ having larger variances in Fig.~\ref{thnum4}(b) and (d) are among the ones with smaller variances in panels (a) and (c). Comparing Fig.~\ref{thnum4}(c) and (d), one also observes that the timescale separation modifies the variance of the oscillators far from the fast component. The theory and the numerical simulations for the two systems with and without timescale separation are confirmed in Fig.~\ref{thnum4}(e)\,. We then move to the case where the noise correlation times are different in each component. In Fig.~\ref{thnum5}\,, we show both limits $\lambda_{N_\mathcal{S}}\tau_{\mathcal{F}}\ll 1$ and $\lambda_2\tau_{\mathcal{S}}\gg 1$ [panels (a), (b) and orange and red data points in panel (e)], and $\lambda_{N_\mathcal{S}}\tau_{\mathcal{F}}\gg 1$ and $\lambda_2\tau_{\mathcal{S}}\ll 1$ [panels (c), (d) and green and blue data points in panel (e)]. Similar amplification of the fluctuations as in the previous case are observed. However, comparing Fig.~\ref{thnum5}(c) and (d), one remarks that some oscillators well connected to the fast component keep rather small variances while others having fewer connections become more vulnerable.

\subsection{Layered networks}
Here, we check the theory when the system is defined on a layered network, i.e. the slow and fast components each corresponds to one layer. We consider the specific setting where both layers have the same network connectivity and the inter-layer coupling is made through single connections between corresponding oscillators in each component (such structure are sometimes called \textit{multiplex}~\cite{New18book}). In Fig.~\ref{thnum3}\,, the numerical simulations for the variance (dots) match the theory Eqs.~(\ref{eq6}) and (\ref{eqapp}) (solid lines) for various correlation times of the noise [homogeneous in panel (a), i.e. $\tau_{\mathcal{F}}=\tau_{\mathcal{S}}$\, and inhomogeneous in panel (b)]. The red, blue and pink data points correspond to the situation where there is a timescale separation between the two components, while for the orange, green and purple ones, there is no timescale separation. Interestingly, one observes that the two different situations produce similar variances for the phases. As predicted in Sec.~\ref{secl}\,, in the limit where $\tau$ is the longest timescale in the system, the variances in the two situations only differ by a constant, which is small when comparing the blue and green data points in Fig.~\ref{thnum3}(a)\,. However, in Fig.~\ref{thnum3}(b) where the noise correlation times are distinct in each component, one sees that the variances with and without timescale separation differ.

\section{Conclusion}\label{sec4}
Physiological systems are composed of a multitude of synchronized dynamical units evolving on various timescales. It is therefore relevant to investigate how these different timescales impact the synchronization dynamics of networked phase oscillators. Here, we considered networks of synchronized phase oscillators where a timescale separation divides the units into a slow and a fast component. Using Mori-Zwanzig formalism, we derived a reduced dynamical system describing the time-evolution of the slow component which we used to assess the resilience of the slow component by calculating the variance of the phase deviations. We obtained a closed-form expression for the variance of each oscillator as a function of the eigenmodes of the reduced Jacobian. Interestingly, noise propagation from the fast to the slow component essentially depends on the mixing of the different eigenmodes. The precise mixing is given by the inter- and intra-component coupling structures. In particular, we showed that oscillators that have a small variance when there is no timescale separation, might have a strongly amplified variance when there is a timescale separation and they have numerous connections to the fast component. Also, for homogeneous long correlation time of the noise, we found that when the fast and slow component are connected over a layered structure, the variance of the oscillators is mostly insensitive to a timescale separation. \\
The theory presented here highlights the importance of timescales to assess the resilience of coupled phase oscillators. Some oscillators that might be the most robust within one ratio of the timescales, might become the most fragile ones for another ratio (see Figs.~\ref{thnum4} and \ref{thnum5}).\\
While the results of this manuscript were obtained for Kuramoto oscillators, they apply more generally to coupled dynamical system evolving close to a stable fixed so that the linear approximation is valid.\\
Future research should consider more than one timescale separation, evaluate the resilience of the fast component, and go beyond the small fluctuation framework presented here.
\section*{Acknowledgments}
We thank Francesco Caravelli for useful discussions.  This work has been supported by the Laboratory Directed Research and Development program of Los Alamos National Laboratory under project numbers 20220797PRD2
and 20220774ER and by U.S. DOE/OE as part of the DOE Advanced Sensor and Data Analytics Program.

%%% If you don't add the figures in the LaTeX files, please upload them when submitting the article.
%%% Frontiers will add the figures at the end of the provisional pdf automatically
%%% The use of LaTeX coding to draw Diagrams/Figures/Structures should be avoided. They should be external callouts including graphics.
\appendix
\section{No timescale separation}\label{app1}
One can also calculate the variance of the oscillators belonging to $\mathcal{S}$ (and also $\mathcal{F}$) when there is no timescale separation. Assuming as previously that the oscillators in $\mathcal{S}$ and $\mathcal{F}$ are subject to noises with correlation times $\tau_{\mathcal{S}}$ and $\tau_{\mathcal{F}}$ respectively, and that the standard deviations are homogeneous, one has,
\begin{align}\label{eqapp}
    \langle x_i^2 \rangle &= \eta_{0}^2\sum_{\alpha,\beta=2}^{N_{\mathcal{S}}+ N_{\mathcal{F}}}\sum_{j\in \mathcal{S}}\frac{(\gamma_\alpha + \gamma_\beta+- 2\tau_{\mathcal{S}}^{-1})q_{\alpha,j}q_{\beta,j}q_{\alpha,i}q_{\beta,i}}{(\gamma_\alpha + \gamma_\beta)(\tau_{\mathcal{S}}^{-1}-\gamma_\alpha)(\tau_{\mathcal{S}}^{-1}-\gamma_\beta)}\\
    &+\eta_{0}^2\sum_{\alpha,\beta=2}^{N_{\mathcal{S}}+ N_{\mathcal{F}}}\sum_{j\in \mathcal{F}}\frac{(\gamma_\alpha + \gamma_\beta - 2\tau_{\mathcal{F}}^{-1})q_{\alpha,j}q_{\beta,j}q_{\alpha,i}q_{\beta,i}}{(\gamma_\alpha + \gamma_\beta)(\tau_{\mathcal{F}}^{-1}-\gamma_\alpha)(\tau_{\mathcal{F}}^{-1}-\gamma_\beta)}\,,\nonumber
\end{align}
where ${\bf q}_\alpha$ are the eigenvectors of the Jacobian Eq.~(\ref{eqjac})\,, with corresponding eigenvalues $\gamma_\alpha$\,.

%\bibliography{bibliography.bib}
%\bibliographystyle{apsrev4-2}

\end{document}